\documentclass[showpacs,prl,superscriptaddress,twocolumn,floatfix,10pt]{revtex4}
\usepackage{graphicx}

\begin{document}

\title{Superconducting diamagnetic fluctuations in ropes of carbon nanotubes}
\author{M.~Ferrier}
\affiliation{Laboratoire de Physique des Solides, associe\'e au CNRS UMR 8502, B\^atiment 510, Universit\'e Paris-Sud, 91405 Orsay, France}
\author{F.~Ladieu}
\author{M.~Ocio}
\affiliation{SPEC CEA 91191 Gif/Yvette France}
\author{B.~Sac\'ep\'e}
\affiliation{SPEC CEA 91191 Gif/Yvette France}
\author{T.~Vaugien}
\affiliation{ Centre de Recherche Paul Pascal/CNRS UPR 8641, Av. du Docteur Schweitzer, 33600 Pessac, France} 
\author{V.~Pichot}
\author{P.~Launois}
\author{H.~Bouchiat}
\affiliation{Laboratoire de Physique des Solides, associe\'e au CNRS UMR 8502, B\^atiment 510, Universit\'e Paris-Sud, 91405 Orsay, France}


\begin{abstract}
We report low-temperature magnetisation measurements on  a large number of purified ropes of  single wall carbon nanotubes. In spite of  a large  superparamagnetic contribution due to the small ferromagnetic  catalytical particles still present in the sample, at  low temperature ($T  < 0.5K$)  and low magnetic field ($H < 80 Oe$), a diamagnetic signal is detectable. This low temperature diamagnetism can be interpreted as the Meissner effect in ropes of carbon nanotubes which have previously been shown  to exhibit superconductivity from transport measurements \cite{kociak01}. 
\end{abstract}

\maketitle

\newpage
  During the last decade, electronic properties of carbon nanotubes have attracted a lot of interest \cite{dressel}. Depending on their diameter and helicity single wall carbon nanotubes  (SWNT) are either semiconducting or metallic with in the latter case  only 2 conducting channels at the Fermi energy. These long molecular wires constitute a model system for the investigation of one dimensional (1D) electronic transport and  its great sensitivity to electron-electron interactions.   
 The usual Fermi liquid is expected to be unstable  in 1D, with the formation of a  correlated Luttinger Liquid state \cite{TLL}. In SWNT,  interactions contain a strong repulsive Coulomb long range contribution,  giving  rise to a characteristic power law depression of the  low energy tunneling density of states whose exponent is related to the interaction strength.  Accordingly, non linear conductance measured on SWNT mounted on tunnel contacts  has been interpreted as the signature of the existence of  a Luttinger  liquid state with repulsive interactions \cite{bocracth}. 
 In this context the discovery of superconductivity below 0.5 K  in transport measurements on SWNT ropes  containing between 30 to 100 parallel 1.2 nm  diameter tubes in good contact with normal electrodes \cite{kociak01,kasu03}  was  a surprise. This observation,  which implies the existence of attractive interactions  overcoming repulsive ones, has stimulated  a number of theoretical investigations \cite{gonzalez,sedeki,demartino02,demartino03,ferrier04} of possible, in most cases phonon mediated, attractive mechanisms. In particular the famous  acoustic breathing modes of SWNT \cite{dressel} have been shown to provide a phonon mediated attractive electron- electron coupling in SWNT when they are assembled  into ropes.  Repulsive Coulomb interactions are  indeed strongly screened by the nearest neighbor metallic tubes, and there is also the possibility of an important inter-tube Josephson coupling \cite{gonzalez,demartino03,ferrier04}. A good agreement between theory and experiment was achieved concerning the broadening  of the transition  observed on the resistance versus temperature  data  when decreasing the  number of tubes constituting the rope.

Beside  transport, it is also essential to investigate thermodynamic signatures such as the Meissner effect of this unusual superconductivity. The geometry of carbon nanotubes,  which  are very narrow 1D cylinders, is a priori  not favorable  for efficient magnetic flux expulsion.  Nevertheless, magnetisation measurements performed on very small  (0.4 nm ) diameter SWNT, revealed a diamagnetic contribution increasing at low temperature  below 6K \cite{tang}  which was interpreted as superconducting fluctuations. However the geometry of the samples grown in zeolite matrices was not adequate for transport measurements, which did not show clear sign of superconductivity.

The aim of the present work is to address this question of flux expulsion (i.e. Meissner effect) in ropes of SWNT similar to those previously studied in transport measurements\cite{kociak01,kasu03}. 

\begin{figure}
\begin{center} 
\includegraphics[width=8cm]{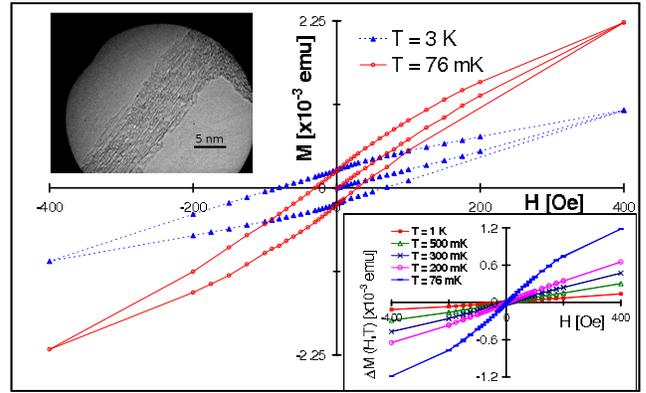}
\end{center}
\caption{Field dependence of the total magnetisation of the sample at various temperatures below 3.5K. Lower inset: field dependence of  $\Delta M(T,H)= M(T,H)-M(3.5, H)$ for different temperatures showing that the hysteretic component of the magnetisation is independent of temperature below 3.5K. (This remains true up to T= 35K.) Upper inset: transmission electron microscopy  image of the SWNT ropes constituting the sample.}
\label{hyst} 
\end{figure}

The samples  consist of  SWNT ropes (containing typically 10 to 100 parallel tubes whose diameter varies between 0.9 to 1.4 nm, the intertube distance being of the order of 0.3nm (see fig.\ref{hyst}). They were fabricated  by  decomposition of Fe(CO)$_5$  according to the HiPCO process \cite{hipco}. Removal of the Fe catalyst was achieved  through annealing under Oxygen at 300 \r{}C and   subsequent treatment with hydrochloric acid. 
After these operations the amount of  nanoparticles of Fe both in reduced or oxidized forms  is estimated to be  of the order of 1$\%$.   These ropes were assembled into long ribbons in a PVA solution using the method described in \cite{poulin}. The  spreading of  alignment of the nanotubes within  a single ribbon was determined from Xray diffraction measurements according to the method described in  \cite{RX}  and found to be $ \pm 35$ \r{}.  For the purpose of this experiment these ribbons were wrapped around two 7 by 7mm high purety  saphyre substrates, which were subsequently annealed 3 hours at 320 \r{}C in order to eliminate the PVA surrounding the ropes. The mass of SWNT ropes  deposited in this procedure, was measured to be 5mg and 6mg on the two saphyre substrate. The  spreading of  alignment of the nanotubes  after wiring on the saphyre is probably  not better than $ \pm 45$ \r{}.  We will see that this nonetheless permits  a determination of the anisotropy of the magnetisation of the nanotubes. 
Magnetisation measurements were performed  on  the 2  samples  in a home built  extraction SQUID magnetometer \cite{ocio} equipped with a dilution  refrigerator insert whose base temperature is 70mK. The thermalisation of the sample was done with a small bundle of high purity copper wires attached to the saphyre plates inserted into a tube made from a capton foil. The contribution of the empty sample holder, capton, saphyre and copper bundle was shown to be below  3 $10^{-8}$emu/Oe  in the whole range of temperature (70mK, 300K) and magnetic field (3 to 3000 0e) investigated.

The magnetic signal contains a large ferromagnetic like hysteretic contribution.   One sees on Fig.\ref{hyst} that this hysteresis   does not change with temperature  from 70mK up  to 3.5K . This remains true up to 35K (not shown). At any  temperature below 35K, we thus  decompose the magnetic signal into a  T independent contribution  attributed to frozen large magnetic particles and  into a temperature dependent part which is  reversible (i.e non hysteretic). In the following we mainly focus on this  reversible low temperature dependent part of the magnetisation  of the samples. We will show that beside the contribution of unfrozen superparamagnetic particles, it is possible to identify the presence of a diamagnetic contribution growing below 0.4 K which will be interpreted as the Meissner effect of the carbon nanotubes.  

\begin{figure}
\begin{center} 
\includegraphics[width=8cm]{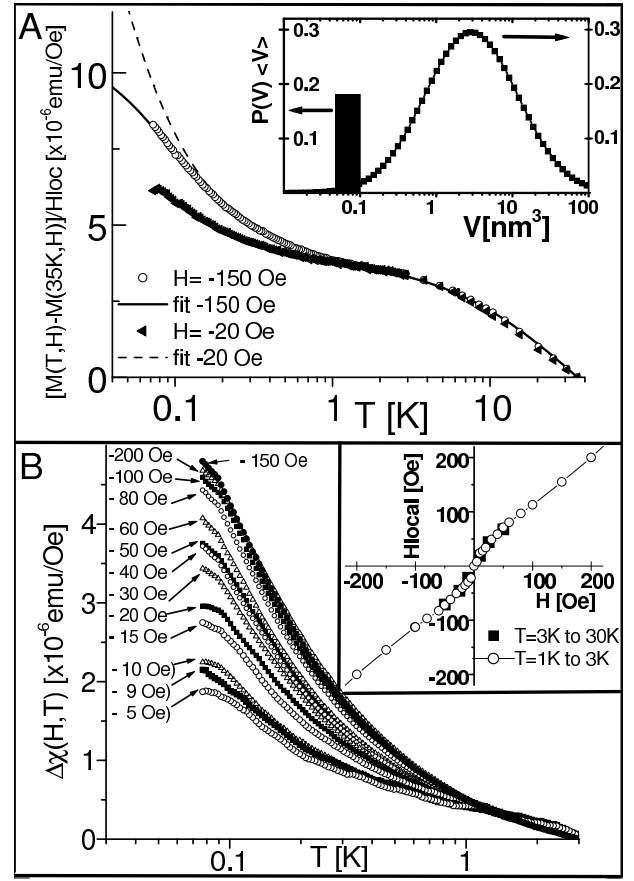}
\end{center}
\caption{ (A)Renormalised magnetisation  $M(T,H) - M(35K,H) /H_{loc}(H)$ as a function of temperature between 70mK and 35K  for H=20 Oe and H=150 Oe compared to the fit with 2 assemblies of superparamagnetic particles whose  adjusted volumic distributions are shown in the inset. Note the deviations  at low temperature for the data taken at 20 Oe.(B)Temperature dependence of the field cooled renormalised magnetisation: $\Delta \chi(T,H)=\Delta M(T,H)/H_{loc}(H)$ for various values of magnetic field. Inset: Field dependence of the local field obtained by renormalisation of the temperature dependent data above 1K up to 3.5K (circles) and up to 30K(squares)}  
\label{mdeT} 
\end{figure}

The magnetisation as function of temperature is plotted in fig.2-A.  $M(T,H)$ increases  monotonously as the temperature is lowered below 35K down to 70mK. This temperature dependent signal  corresponds approximatively to 50\%  of the total magnetisation of the sample. The data was systematically taken in the field cooled state  \cite{FC}.  It is possible to superimpose  {\it all} the  curves  taken at various values of magnetic field below 150 Oe and above 1K by  renormalisation  of the data with  a field dependent factor $H_{local}(H)$ which corresponds to the effective average  magnetic field inside the sample. As shown  in fig.2-B $H_{local}(H)$ deviates from the applied magnetic field at low field values which suggests that the actual field inside the sample has a contribution  coming from the largest particles saturating at low  fields  (typically lower than 30 Oe) which adds to the  applied field. We have checked that the determination of $H_{local}(H)$ does {\it not} depend on the temperature  range chosen between 1 and 35K to renormalise the curves. More over  in this temperature range, as shown on the inset of fig.2A, the quantity $\ M(T,H_{local}(H))$ can be very well described within a simple model of a   volumic distribution of superparamagnetic $Fe_2O_3$ oxyde particles  as inspired from previous work in ref. \cite{sappey}  where the anisotropy energy  of $Fe_2O_3$ is taken to be K = 5.4 $10^4 J/m^3$. This  distribution shown in the inset of fig.2-A, has a broad peak  characteristic of a lognormal distribution 
 around a particle volume which corresponds to typical diameters of the order of 1.5nm, most of these particles are frozen below 3-10K.  To account of the very low T upturn of M(T) we have added to the previous contribution a Brillouin law with J=8. This    corresponds to  very small $Fe_2O_3$  aggregates containing typically 3 or 4  coupled iron spins. The magnetic moments of these particles (whose anisotropy energy is estimated to be of the order of 20mK) are unfrozen down to very low temperature.
 Finally, the amplitudes of these 2 superparamagnetic signals  added to the ferromagnetic part (independent of temperature)   corresponds to a number of typically 3 $10^{17}$ Fe atoms  for a 5 mg sample i.e 1\% of the sample. This is what is expected after purification, in agreement with previous magnetisation measurements made on similar samples \cite{mcatal} above 4K. On can see that the magnetisation signal at 150 Oe is very well described
by this simple model between 60K and 70mK \cite{hightemp}. On the other hand important  low T deviations from this model are observed for the data measured at 20 Oe, increasing  below 0.5K. 

\begin{figure}
\begin{center} 
\includegraphics[width=8cm]{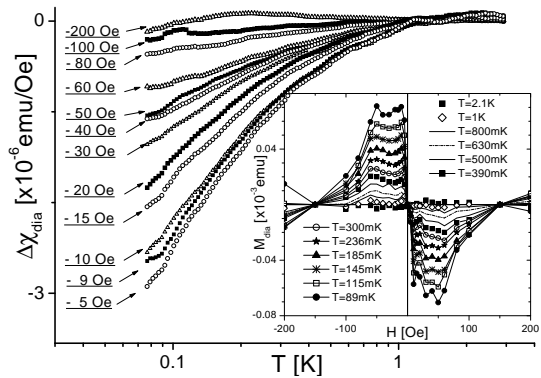}
\end{center}
\caption{Temperature dependence of the renormalised magnetisation after substraction  of  the data at $H=150 Oe$:  $\Delta \chi _{dia}(T,H) = \Delta \chi(T,H) - \Delta \chi(T,150). $Inset: Reconstitution,  using  eq.4, of the field dependence of the diamagnetic contribution for various temperatures.} 
\label{mdia} 
\end{figure}

   If we now focus on this low temperature data below 0.5 K, we can see on  fig.2-B, that the  temperature dependence of  $\Delta \chi(T,H) = \Delta M(T,H) /H_{local}(H)$   with $\Delta M(T,H)=(M(T,H) -M(T=3.5K))$ varies substantially  with the  magnetic field.  Note that  the magnetisation of an assembly of  {\it unfrozen} super-paramagnetic   particles  of typical moment $\mu$ is indeed expected to exhibit at low  temperature a  non linear field dependence   starting in the field range where $\mu H \sim k_BT$. This consideration can explain the data above 150 Oe quite well, where a decrease of the T dependence of  $\Delta \chi$  with magnetic field is indeed observed (see the curve at 200 0e in fig.2).  
On the other hand the low field  data  is surprising.   Below 1K  $\Delta \chi$ is observed  (see fig.2) to {\it strongly  decrease } with decreasing magnetic field between 100 and 5 Oe.  Such a behavior {\it cannot} correspond to any superparamagnetic system even in the presence of interactions between the particles leading to a spin-glass like state \cite{SG}. This effect,  is emphasized in fig.\ref{mdia} showing the quantity $\Delta \chi_{dia}(T,H)$ after performing the subtraction:

\begin{equation}
\Delta \chi_{dia}(T,H) = \Delta \chi(T,H) - \Delta \chi(T,150)
\end{equation}

This plot strongly suggests the existence of a  temperature dependent diamagnetic signal increasing  rather smoothly at  temperatures below 0.4K, for low magnetic field, superimposed to the superparamagnetic contribution of the catalyst particles. By subtracting data at 150 Oe instead of the superparamagnetic contribution inferred from the fit of fig.2-A, we probably underestimate this diamagnetic signal, on the other hand this method is model independent.

 In the following we show that this diamagnetic contribution  can be interpreted as  screening diamagnetic supercurrents  due to superconducting fluctuations in the ropes of carbon nanotubes.   These ropes can be approximated by superconducting  cylinders  with diameter D  small compared to the  London penetration  length $\lambda(T)$  (see for example the recent work on  Pb superconducting  nano-cylinders \cite{supnanopb}). In the linear response regime,(observed in present case  up to 10 Oe) the diamagnetic susceptibility (per unit volume) of a  dense assembly of  superconducting cylinders  can be simply expressed as a function of $D/\lambda$ as:
 
\begin{equation}
 M_{\|,\bot}(H) = -H (D/\lambda_{\|,\bot}(T))^2
\end{equation}
  
  when the field is respectively parallel, perpendicular to the axis of the nanotubes within the rope.
   $\lambda_{\|}(T)$and $\lambda_{\bot}(T)$ are respectively the penetration length  parallel and perpendicular to the cylinders. Considering
the  anisotropy of the electronic structure of a single tube,  as well as the existence of an inter-tube Josephson coupling, it is reasonable to assume that  $\lambda_{\|}(T)$ and $\lambda_{\bot}(T)$ may be different from one another.  In the absence   of a more  precise theoretical prediction, we use the simple London expression:   $\lambda_{\| , \bot}^{-2}(T)\simeq 2 \mu_0 n_S(T) e^2/m_{\bot ,\| }$ where $ n_S (T)$ is the density of superconducting pairs in the ropes and $m_{ \bot, \|}$ are respectively the effective masses characterizing  transport  parallel  and perpendicular to the ropes.  Since only 1/3 of the tubes within a rope are expected to be metallic, we assume that at zero temperature $n_s(0) = n_e/6$, where   $n_e = \frac{4}{ a \pi d^2/4}$  is the density  of electrons in the 2 conduction bands of a metallic SWNT of diameter $d$ and $a$ is the graphene hexagonal lattice size. As a result, we get a simple expression relating the London diamagnetic susceptibility   to the geometry of the carbon nanotubes ropes assumed to take random orientations through the sample (anisotropy effects will be discussed later).
\begin{equation}
  M_{av}^{dia}(H) = - H\frac {8 e^2}{3 m a \pi}\frac{D^2 }{d^2}
  \label{chidia}
  \end{equation}
(m is approximated here by the free electron mass).
 From our measurements we can deduce the amplitude of the diamagnetic susceptibility in the linear regime  to be $\chi_{meas}^{dia} = 3\pm 0.05$    $10^{-6}$ emu/Oe $(A m ^2/T)$,  reasonably  close to  the value  given by  expression  (\ref{chidia}) (multiplied by the volume of the sample), $\chi_{cal}^{dia} = 2$ $10^{-5} emu/Oe$ for an average diameter $D= 10 nm$ per rope  as estimated from transmission microscopy. This yields $\lambda(0) \simeq 0.6\mu m  $ indeed much longer than the average diameter of the ropes. The temperature below which diamagnetism shows up ($~ T^* =0.4K$) is also very close to the values of transition temperature measured in transport measurements \cite{kociak01}.

From the set of  curves  in fig.\ref{mdia} it is possible to reconstruct the field dependence of this diamagnetic component of the magnetisation for various temperatures, as shown in the inset:

  \begin{equation}
 M_{dia}(H) =\Delta M(H) - H_{local}(H)\Delta \chi(150)
  \end{equation}
  
$M_{dia}(H)$  measured at  70mK varies linearly at low field and goes through a maximum  around 40 Oe, it   decreases at higher field  and becomes undetectable above 80 Oe. This behavior observed on both samples is  typical of the magnetisation curves of a type 2 superconductor  whose  critical field $H_{c1}$ is of the order of $60  \pm 20$ Oe.
Note   that  this diamagnetic contribution at the lowest investigated  values of magnetic field and temperature constitute  more than 2/3 of the  temperature dependent magnetic signal  below 3K and 1/3 of the total magnetic contribution.

We have also compared the  amplitude of the magnetisation when the magnetic field is parallel to the directions of wrapping of the fibers compared to the situation where it is perpendicular. In both cases the field was maintained in the plane of the saphyre plate in order to avoid any macroscopic demagnetising factor. The paramagnetic component of the sample is found to be isotropic within an accuracy of 1\%. On the other hand it was possible to detect unambiguously  on both samples an { \it anisotropy} of the Meissner component of the signal of the order of 10\%$\pm 1$\% indicating a larger amplitude of the Meissner component along the axis of the tubes than perpendicularly by  a factor 1.2$\pm 0.05$  taking an angular distribution of the tubes axis  of  $\pm$45\r{}   wide, (note that this number only slightly depends on the estimated spreading angle when its smaller than 60\r{}). This result is however surprising since the Josephson  coupling energy $t_j$ between the tubes is expected to be much smaller (at least by a facter 100) than the hopping integral $t$ along the tubes axis \cite{gonzalez}.  Nevertheless it is  possible that the effective mass which characterize the intertube hopping of the Cooper pairs $m_{\bot}\propto 1/d^2 t_j$,  is of the order and even  smaller than along the tube axis $m_{\|}\propto 1/a^2 t$ since $d/a \simeq 10$.

All these findings corroborate the   superconductivity observed in transport measurements on individual ropes of carbon nanotubes. Note  that the diamagnetic part of the magnetisation disappears at relatively low magnetic field, more than a factor 10 lower than the critical field of the order of 1T observed in transport experiments. This can be understood  considering that the weak Josephson  energy coupling between the tubes within one rope  is destroyed at much weaker magnetic field  than superconductivity within a single tube giving rise to a   diamagnetic signal per unit volume of the order of $d^2/ \lambda ^2$ (undetectable in this experiment).

In all this analysis we have not taken into account  the contribution of the orbital magnetism of the 
tubes in the normal state. Such orbital currents have been predicted  and measured in multiwall nanotubes \cite {magMWNT}. They are related  to the sensitivity of electronic eigenstates and energies to the phase 
of the boundary conditions modified by the magnetic field.  The resulting magnetic orbital susceptibility is  diamagnetic for semiconducting tubes and paramagnetic for metallic tubes \cite{liu} and can be estimated to be at most $ \chi_{orb}= 2 $ $10 ^{-8} emu  Oe^{-1}$  for  our sample, which is negligible compared to the diamagnetic contribution we have discussed.     
Our findings also present qualitative similarities with the results obtained by Tang et al. \cite{tang} on 0.4 nm diameter tubes, except that temperature and field scales are much smaller in our experiment. This could simply be related to the diameter of the tubes which are in the present case 3 times larger than in the Tang experiment. The characteristic energy of the phonon breathing mode, expected to be at the origin of superconductivity \cite{demartino02,demartino03} is known to scale as the inverse tube diameter. Moreover the possibility of the existence of a  Wentzel Bardeen  instability \cite{martin} specific to electron phonon coupling in a 1D system is  predicted in very small diameter tubes, \cite{demartino02} favoring superconducting fluctuations against the formation of a Luttinger liquid with repulsive interactions, even in the situation where they are not well screened.  
 Finally we note that the Kondo effect,  leading to a dynamical screening   of the magnetic moments at low temperature, could also give rise to a low temperature decrease of the magnetic susceptibility as observed in noble metals with a low concentration of magnetic impurities. To our knowledge there is no estimate of the order of magnitude of the Kondo temperature of iron magnetic impurities in carbon nanotubes. Since  the electronic density of states is very small in carbon nanotubes compared to  standard metals we expect this Kondo temperature  to be also extremely small.  Even if such a Kondo physics scenario cannot completely be ruled out  to explain  the depression of the Curie law at low field and low temperature it could not provide a description of the field dependence of the data shown in fig.\ref{mdia}.
 
In conclusion, we have shown that magnetisation of ropes of SWNT,  at low temperature and magnetic field,  cannot be explained by the superparamagnetism  of the catalytical particles present in the sample.  The weaker relative increase of magnetisation  at  small magnetic field compared to higher one, provides a strong indication of the existence of a diamagnetic Meissner contribution in  ropes of SWNT increasing below 0.4 K, in agreement with the onset of superconductivity observed in  transport measurements realised on similar samples.  From the   alignment of the tubes in  the samples, although imperfect, it was also possible to  probe the anisotropy of the Meissner effect.

We thank  S.Gueron, R.Deblock, M.Kociak, P.Poulin, S.Senoussi, L.Lepape, E.Vincent, J.Hamman and P.Bonville  for fruitful discussions and help.


\begin{thebibliography}{99}
\bibitem{FC}A small albeit systematic difference of the order of few  $\%$ of 
$\Delta M(H,T)$ at 100mK  could be noticed between the field cooled and zero field cooled data. We  chose to focus on the FC data since it yields the under bound of $\vert M_{dia} \vert$.
\bibitem{hightemp}Deviations  to the fit observed at temperatures above 100K can be attributed to the largest particles whose  magnetisation freeze out in this temperature range  contributing to the local field and not taken into account by the 2 distributions fit we have used.
\bibitem{SG} In spin glasses at any given T, M(H)/H is a decreasing function of H.
\end{thebibliography}
\end{document}